\documentclass{article}
\usepackage{spconf,amsmath,graphicx}

\usepackage{subcaption}
\usepackage{hyperref}

\title{Room Impulse Responses help attackers to evade Deep Fake Detection}
\name{Hieu-Thi Luong\textsuperscript{1}\sthanks{Dr. Hieu-Thi Luong is funded by RIE2025 NRF International Partnership Funding Initiative. This research is supported by the National Research Foundation, Singapore, under the AI Singapore Programme (AISG Award No.: AISG2-TC-2023-011-SGIL). Any opinions, findings and conclusions or recommendations expressed in this material are those of the author(s) and do not reflect the views of National Research Foundation, Singapore.}, Duc-Tuan Truong\textsuperscript{1}, Kong Aik Lee\textsuperscript{2}, Eng Siong Chng\textsuperscript{1}}
\address{\textsuperscript{1}Nanyang Technological University, Singapore \\\textsuperscript{2}The Hong Kong Polytechnic
University, Hong Kong SAR, China}
\begin{document}
%\ninept
%
\maketitle
\begin{abstract}
The ASVspoof 2021 benchmark, a widely-used evaluation framework for anti-spoofing, consists of two subsets: Logical Access (LA) and Deepfake (DF), featuring samples with varied coding characteristics and compression artifacts.
Notably, the current state-of-the-art (SOTA) system boasts impressive performance, achieving an Equal Error Rate (EER) of 0.87\% on the LA subset and 2.58\% on the DF.
However, benchmark accuracy is no guarantee of robustness in real-world scenarios.
This paper investigates the effectiveness of utilizing room impulse responses (RIRs) to enhance fake speech and increase their likelihood of evading fake speech detection systems. Our findings reveal that this simple approach significantly improves the evasion rate, doubling the SOTA system's EER.
To counter this type of attack, We augmented training data with a large-scale synthetic/simulated RIR dataset.
The results demonstrate significant improvement on both reverberated fake speech and original samples, reducing DF task EER to 2.13\%.
\end{abstract}
\section{Introduction}
\label{sec:intro}

The artificial intelligence (AI) boom has increased the accessibility of various state-of-the-art (SOTA) generative AI tools, which can synthesize text \cite{floridi2020gpt}, image \cite{rombach2022high}, video \cite{thies2016face2face,chan2019everybody}, audio \cite{kreuk2022audiogen}, or speech \cite{arik2018neural} that are indistinguishable from authentic media \cite{masood2023deepfakes}.
The Automatic Speaker Verification Spoofing and Countermeasure (ASVspoof) Challenge addressed the problem of detecting spoofed speech and audio deepfakes \cite{kinnunen2017asvspoof,todisco2019asvspoof}. The last edition, ASVspoof 2021 \cite{yamagishi2021asvspoof}, focused on differentiate between the real (bonafides) and fake (spoofed) speech in the Logical Access (LA) and Deepfake (DF) scenarios. The spoofed audios used in these tasks were constructed using different types of text-to-speech (TTS) \cite{casanova2022yourtts,shen2023naturalspeech} and Voice Conversion (VC) systems \cite{vyas2023audiobox}.
Muller et\ al. \cite{muller2024new} contend that binary classification of fake and real speech is overly simplistic, as audio manipulation encompasses a broader range of techniques beyond speech synthesis, including spatial and environmental editing.
In a more sophisticated attack scenario, adversaries can leverage room impulse responses (RIRs) to inject reverberation into fake speech, effectively mimicking real-world acoustic environments. This can be achieved by estimating RIRs from reference audios \cite{steinmetz2021filtered}, images \cite{singh2021image2reverb}, or videos \cite{ratnarajah2023av}.
These schemes escalate the fight against fake speech, making detection increasingly challenging.

In this research, we investigated the dual role of RIRs in fake speech detection, examining both their potential to compromise system security and enhance its robustness.
We first showed how attackers can bypass detection systems using reverberant fake speech, then evaluated its performance when using RIR-augmented training datasets.
Our results reveals that augmenting with large-scale synthetic RIRs not only substantially enhanced detection rates for reverberant fake speech but also appeared to improve the model's generalizability across various encoding and compression methods.
Building on existing research that utilizes simulated RIRs for various speech \cite{ko2017study} and audio tasks \cite{salamon2017deep,bryan2020impulse}, we extended this concept to the domain of fake speech detection, from both adversarial and defensive perspectives.

This work pursues three objectives: (i)\ Demonstrating the vulnerability of fake speech detection systems to reverberant synthetic speech; (ii)\ Introducing a framework to assess the susceptibility of detection models to specific RIR conditions; (iii)\ Verifying that augmenting training datasets with large-scale synthetic or simulated RIRs enhances model robustness.
This paper is organized as follows: Section \ref{sec:rir} presents our methodology for synthesizing RIRs; Section \ref{sec:scenarios} explores two scenarios where RIRs can be leveraged to compromise or enhance detection systems; Section \ref{sec:experiments} outlines our experimental design and results; Section \ref{sec:discussion} examines key observations and implications; Section \ref{sec:conclusion} summarizes our findings.

\section{Room Impulse Response Synthesizing}
\label{sec:rir}

RIRs are typically characterized by two parameters: reverberation time (T60) and Direct-to-Reverberant Ratio (DRR). In this work, 
we complemented prerecorded RIRs with a synthesizing method based on Bryan's technique \cite{bryan2020impulse}, which controls DRR and T60 characteristics of synthetic signals.
Specifically, an RIR, denoted as $h(t)$, can be decomposed into two distinct components: the early response, $h_e(t)$, and the late-field response, $h_l(t)$:
\begin{align}
    h_e(t) &= \begin{cases} h(t) & t_d-t_0 \leq t \leq t_d+t_0  \\ 0 & \text{otherwise} \end{cases} \\
    h_l(t) &= \begin{cases} h(t) & t > t_d+t_0 \\ 0 & \text{otherwise} \end{cases}
\end{align}
where $t_d$ represents the time delay of the direct path, and $t_0$ denotes the tolerance window, which was set at 2.5ms.
We generated a synthetic RIR with the desired DRR and T60 by manipulating the early and late-field responses respectively \cite{bryan2020impulse}.
The same recorded RIR can be reused to generate multiple synthetic ones with distinct properties. For detailed instructions, readers are referred to the original paper \cite{bryan2020impulse}.

\begin{figure}[t]
  \centering
  \includegraphics[width=\linewidth]{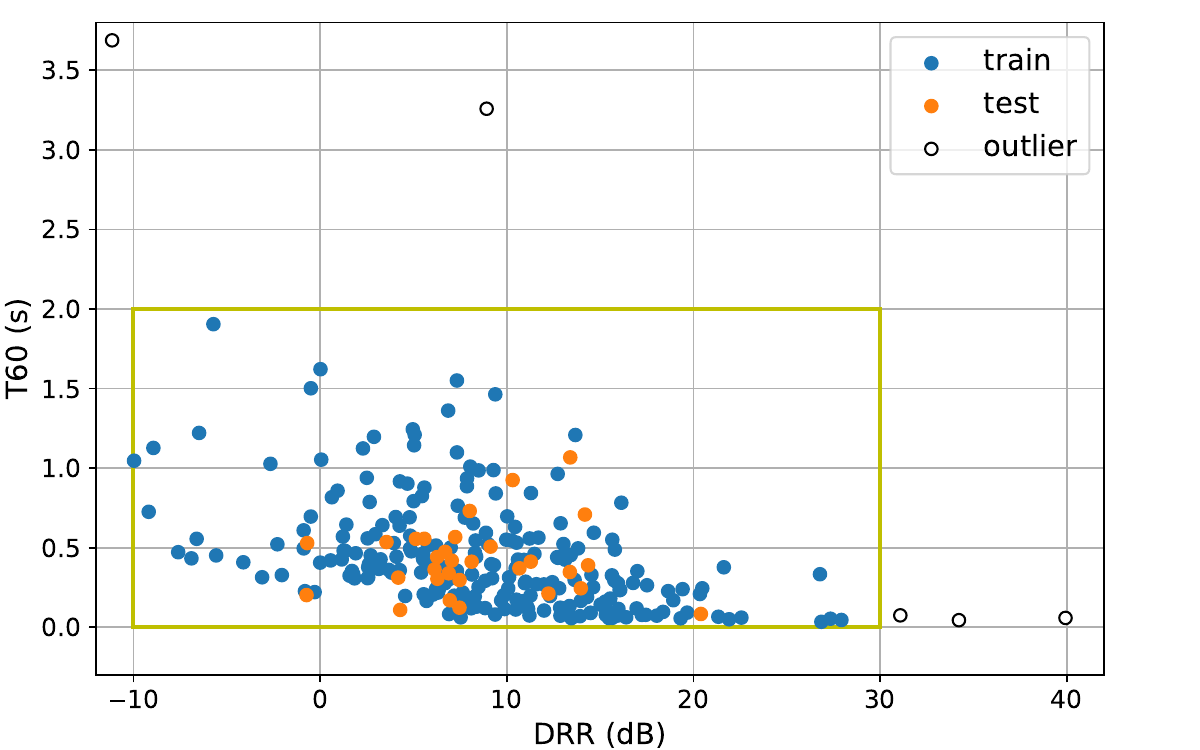}
  \caption{Scatter plot of T60 and DRR for MIT RIRs.}
  \label{fig:recorded_irs}
\end{figure}
\begin{figure}[t]
  \centering
  \includegraphics[width=\linewidth]{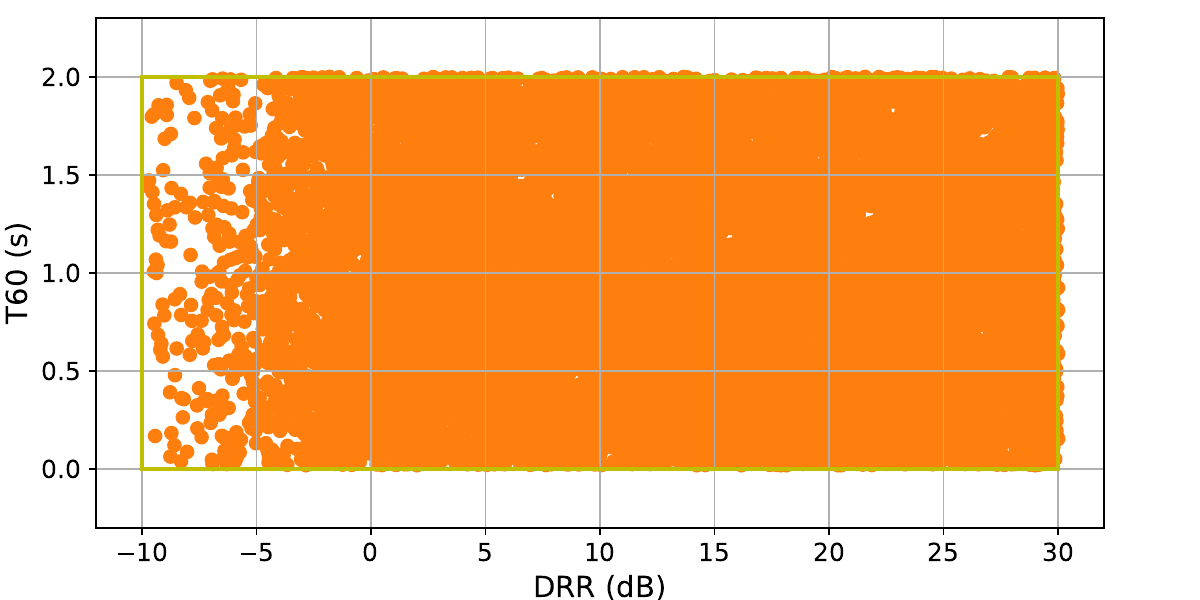}
  \caption{Scatter plot of T60 and DRR for the synthetic RIR test set.}
  \label{fig:synthetic_ir_test}
  \vspace{-3mm}
\end{figure}

For the recorded RIRs, we utilized the MIT RIR dataset published by Traer and McDermott \cite{traer2016statistics} in 2016, comprising 270 impulse responses captured in diverse environmental settings.
Figure \ref{fig:recorded_irs} illustrates the distribution of T60 and DRR values within the MIT RIR dataset.
We used the same method of estimating T60 and DRR for both calculating and synthesizing as described in \cite{bryan2020impulse}.
For the experiment, we excluded RIRs with T60 exceeding 2.0 seconds or DRR outside the range of -10 to 30 dB.
From the remaining, we randomly selected 30 samples to create the test set (ir-test-real) and designated the rest for training (ir-train-real).

We created the synthetic RIR datasets by passing each recorded RIR from the test and train sets through the synthesizing process 500 times, resulting in a synthetic test set (ir-test-syn) with 15,000 samples and a synthetic train set (ir-train-syn) with 117,000 samples.
The T60 and DRR values were uniformly distributed across the range of 0.02 to 2.0 seconds and -10 to 30 dB respectively.
However, when performing DRR augmentation, we rejected any combination where the late-field maximum exceeded the early response and attempted a new combination instead.
Figure \ref{fig:synthetic_ir_test} illustrates the distributions of T60 and DRR values in the synthetic test set obtained through our process.
Interestingly, RIRs with DRR values between -10 and -5 dB exhibited a significantly higher rejection rate compared to those outside this range.
A similar trend was also evident in the synthetic RIR train set.

In addition to synthetic RIRs, we also utilized a simulated RIR dataset\footnote{https://www.openslr.org/26/}, comprising 60,000 samples simulated using the image method \cite{allen1979image}. 
The dataset was published by Ko et.\ al \cite{ko2017study} as part of their research on reverberant speech augmentation for robust speech recognition.
We only used it for training (ir-train-sim) to evaluate and compare the effectiveness of the synthetic and simulated RIRs as training augmentation method.
The primary difference between the two types of RIRs lies in their origin: synthetic RIRs are generated from real-world recordings, whereas simulated RIRs are created using purely mathematical models.

\section{RIRs for Attacking and Defending Fake Speech Detection Systems}
\label{sec:scenarios}

In this work, we exclusively focus on fake speech, specifically synthetic speech generated by machines through TTS or VC systems \cite{yamagishi2021asvspoof}, as opposed to various forms of spoofed speech, which may involves recordings physically collected from a target speaker.
These synthetic speech samples can be used to spread disinformation, falsify recordings, or scam individuals.
From the attacker's perspective, the goal is twofold: first, to evade detection by an automatic fake speech detection system, and then to deceive a human target.

\begin{figure*}[t]
    \begin{subfigure}[b]{0.3\textwidth}
         \centering
         \includegraphics[width=\textwidth]{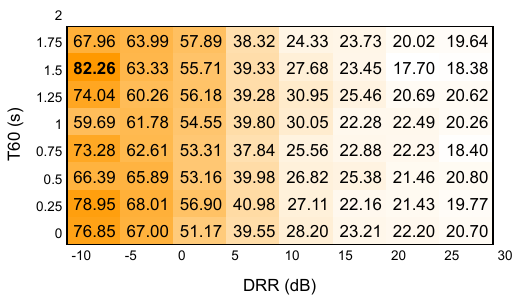}
         \caption{RawNet2}
     \end{subfigure}
     \hfill
     \begin{subfigure}[b]{0.3\textwidth}
         \centering
         \includegraphics[width=\textwidth]{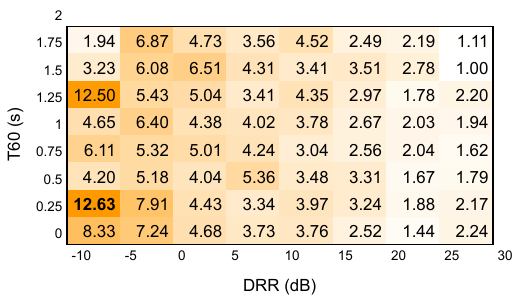}
         \caption{Wav2Vec2 + AASIST}
     \end{subfigure}
     \hfill
     \begin{subfigure}[b]{0.3\textwidth}
         \centering
         \includegraphics[width=\textwidth]{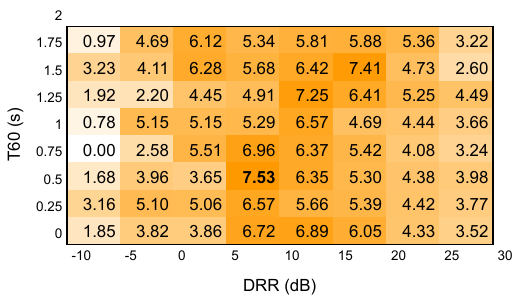}
         \caption{Wav2Vec2 + Conformer}
     \end{subfigure}
     \vspace{-2mm}
        \caption{FAR (\%) of the baseline systems (A, B, and C) at the EER threshold tested on the reverberant C1R2 evaluation set, results split by T60 and DRR values of the augmented RIRs.}
        \label{fig:far-baseline}
  \centering
  \vspace{-2mm}
\end{figure*}

\subsection{For Attackers}
Based on this premise, incorporating RIRs to add reverberation to fake speech could be an effective method to evade detection systems. Although reverberation typically distorts and degrades speech quality, this degradation can be beneficial when attempting to deceive a human listener.
From this point of view, reverberant speech can be considered a type of adversarial sample \cite{carlini2018audio,xie2021enabling}, as the perceived content remains largely unchanged for humans, yet it is intended to alter the classification system's output.
To test this hypothesis, we created two new evaluation sets of reverberant fake speech using recorded and synthetic RIRs, then assessed the impact of reverberation on the performance of detection systems.

\subsection{For Defenders}
Previous works on fake speech detection have explored the use of RIRs for data augmentation \cite{chen2021pindrop,tomilov21asvspoof}. However, these studies relied on conventional public RIR datasets without investigating the potential benefits of synthetic RIRs.
In this paper, we test the effectiveness of using RIR synthesizing method to create large-scale dataset for data augmentation without the need for extensive and costly RIR collection efforts.
In our experiments, we investigated the effect of training fake speech detection systems using three different RIR datasets: small-scale recorded RIRs, large-scale synthetic RIRs, and large-scale publicly accessible simulated RIRs. We evaluated the performance of these systems on various test datasets and reported the results.
Furthermore, we delve into the detailed results to determine whether these systems exhibit increased vulnerability to specific RIR conditions.

\section{Experiments}
\label{sec:experiments}

\subsection{Datasets}
We tested the effect of RIRs on fake speech detection systems by using the ASVspoof 2021 DF subset \cite{yamagishi2021asvspoof} for evaluation and the ASVspoof 2019 LA subset \cite{todisco2019asvspoof} for training.
Performance was evaluated using Equal Error Rate (EER), where the false rejection rate (FRR) and false acceptance rate (FAR) are equal, following the common experimental setup.

The DF subset contains nine conditions marked from C1 to C9, each has 67,981 samples. C1 is the original condition without any codec artifact while the rest are under different lossy compression conditions.
To investigate the impact of RIRs, we generated two additional conditions by applying reverberation to the samples from the original C1 condition, resulting in C1R1 (using ir-test-real) and C1R2 (using ir-test-syn).
Note that we exclusively applied RIRs to the fake speech samples in C1, omitting the bonafide samples, as our goal from the attacker's perspective is to increase the FAR of fake/synthetic audio, making it more challenging for the system to detect.
For comprehensive evaluation, we assessed all fake speech detection systems on four test sets: C1R1, C1R2, C1, and the pooled results of C1 to C9, ensuring that the proposed systems achieved high accuracy on both the new reverberant test data and the legacy DF test set.

Examples of reverberant fake speech can be accessed on our website\footnote{https://hieuthi.github.io/sample-rir-deepfake-detection}.
We invite readers to listen to them and observe how the addition of reverberation can mask artifacts and imperfections in machine-generated speech, making it more challenging to distinguish from natural speech.

\subsection{Evaluated Systems}
As this study concentrated on investigating the impact of recorded and synthetic RIR augmentations, we did not develop a novel fake speech detection model. Instead, we used three SOTA open-source systems and assessed their performance on both the new reverberant evaluation sets and the RIR augmentation training method, to gauge their robustness and adaptability.
The first system, denoted as \texttt{System A}\footnote{https://github.com/TakHemlata/RawBoost-antispoofing}, is the RawNet2 \cite{tak2021endtoend} trained with the RawBoost data augmentation technique \cite{tak2022rawboost}. Notably, RawNet2 served as one of the baseline systems in the ASVspoof 2021 challenge.
The second system, referred to as \texttt{System B}\footnote{https://github.com/TakHemlata/SSL\_Anti-spoofing}, is another detection system proposed by Tak et\ al. \cite{tak2022automatic}. It leverages a pretrained Wav2Vec 2.0 model \cite{baevski2020wav2vec} and incorporates a spectro-temporal graph attention network (AASIST) \cite{jung2022aasist}, achieving high accuracy on both tasks of the ASVspoof 2021 challenge.
The third system, \texttt{System C}\footnote{https://github.com/ErosRos/conformer-based-classifier-for-anti-spoofing}, is a Wav2Vec-2.0-based approach proposed by Rosello et\ al. \cite{rosello2023conformer}. It employs a Conformer model \cite{gulati2020conformer} for classification and achieves SOTA performance on the DF task at the time of writing, surpassing existing benchmarks.
The authors of these works made their official source code publicly available, which we used for our experiments.

In addition to the three baseline systems (\texttt{A},  \texttt{B}, and  \texttt{C}) retained from their original papers, we modified their source codes to accommodate RIR augmentation training.
We then trained three additional sets of systems: The first set (\texttt{AR1}, \texttt{BR1}, and \texttt{CR1}) was trained using speech augmented with recorded RIRs (ir-train-real); The second set (\texttt{AR2},  \texttt{BR2}, and \texttt{CR2}) utilized synthetic RIRs (ir-train-syn), which was 500 times larger than the recorded RIR set; The third set (\texttt{AR3},  \texttt{BR3}, and \texttt{CR3}) employed simulated RIRs (ir-train-sim), approximately half the size of the synthetic RIR set.

\subsection{Training setup with RIR augmentation}
Unless otherwise stated, we retained the original training configurations for each system. All three systems used RawBoost with stationary noise, as previously reported to achieve optimal DF evaluation set performance \cite{tak2022automatic}. We preserved each system's unique training processes and hyperparameters, only modifying them to accommodate RIR augmentation training.
During RIR augmentation training, we added reverberation to training samples by randomly selecting an RIR from the inventory 99\% of the time.
This approach allowed us to increase the diversity of the training samples while retaining some original data to prevent overfitting.
By conducting reverberation online during training, we created an virtually limitless array of combinations.
The samples were additionally subjected to random scaling between 0.4 and 1, further reducing clipping and increasing diversity.

\begin{table}[tb]
    \caption{EER (\%) on C1, the DF evaluation set (pooled), and the reverberant fake speech C1R1 and C1R2.}
    \label{tab:eer}
    \centering
    \scalebox{0.88}{
    \begin{tabular}{l|rr|rr}
        \hline \hline
        System & \multicolumn{2}{c|}{DF} & \multicolumn{2}{r}{C1 with RIRs} \\
         & C1 & Pooled & C1R1 & C1R2 \\ \hline
        \texttt{A} : RawNet2 & 26.38 & 22.75 & 39.53 & 34.6 \\
        \texttt{B} : Wav2Vec2 + AASIST & \textbf{2.34} & 2.85 & \textbf{4.00} & \textbf{3.54} \\
        \texttt{C} : Wav2Vec2 + Conformer & 2.78 & \textbf{2.58} & 5.90 & 5.10 \\ \hline
        \texttt{AR1} : A w/ ir-train-real & 26.31 & 22.52 & 30.62 & 28.20 \\
        \texttt{BR1} : B w/ ir-train-real & 3.46 & 3.54 & 8.30 & 7.88 \\
        \texttt{CR1} : C w/ ir-train-real & \textbf{2.68} & \textbf{2.52} & \textbf{4.00} & \textbf{3.90} \\ \hline
        \texttt{AR2} : A w/ ir-train-syn & 28.63 & 23.12 & 29.07 & 29.24 \\
        \texttt{BR2} : B w/ ir-train-syn & \textbf{1.72} & \textbf{2.13} & \textbf{2.53} & \textbf{2.57} \\
        \texttt{CR2} : C w/ ir-train-syn & 2.34 & 2.39 & 4.67 & 4.23 \\ \hline
        \texttt{AR3} : A w/ ir-train-sim & 28.03 & 23.20 & 31.48 & 30.28 \\ 
        \texttt{BR3} : B w/ ir-train-sim & 4.07 & 4.33 & \textbf{4.07} & 4.77 \\
        \texttt{CR3} : C w/ ir-train-sim & \textbf{2.51} & \textbf{2.43} & 4.60 & \textbf{4.15} \\ \hline 
        \hline
    \end{tabular}
    }
\end{table}

\subsection{Results}

Table\ \ref{tab:eer} presents the EERs of all detection systems on the evaluation sets.
The first three rows display the results of the baselines, revealing \texttt{System B} outperforms  \texttt{System C} on C1, but this trend is reversed in the pooled results.
As anticipated, the EERs increase substantially when the systems are tested on reverberant fake speech, specifically with C1R1, which shows a significant improvement in bypassing detection systems, yielding a relative EER increase of 70.94\% for  \texttt{System B} and 112.23\% for  \texttt{System C}.
Overall, simply replacing fake samples with their reverberant counterparts can enhance the success rate of evading detection systems, effectively doubling it, with the use of recorded RIRs had a more profound impact than synthetic ones (12.71\% higher in case of  \texttt{System B}).
This suggests that recorded RIRs can create more realistic reverberant conditions, leading to a greater challenge for detection systems.

The subsequent three rows present the EERs of systems trained with recorded RIRs, revealing varying outcomes. Notably, the augmentation proves beneficial for \texttt{AR1} and \texttt{CR1}, consistently improving their performance across all evaluation sets, but it appears to degrade the performance of  \texttt{BR1}.
One possible explanation is that \texttt{BR1} was overfitted to a limited set of RIRs, leading to increased EER. Meanwhile,  \texttt{CR1} consistently outperforms  \texttt{System C} across all evaluation sets, not just those with reverberant conditions, suggesting a more robust performance improvement.
This result indicates that RIR augmentation can be beneficial for training robust detection systems, but its effectiveness is contingent on a diverse set of RIRs, as limited RIRs can lead to overfitting and diminished performance.

The systems trained with either synthetic or simulated RIRs also yield varying results, but generally demonstrate beneficial performance improvements in most cases.
Notably,  \texttt{BR2} achieves a SOTA performance across all evaluation sets, surpassing the previous best results. Specifically,  \texttt{BR2}'s EER is 17.44\% lower than  \texttt{System C}'s on the pooled result, and it attains the lowest EERs on reverberant fake speech among all evaluated systems, demonstrating exceptional performance in this challenging condition.
The differences between systems trained with synthetic and simulated RIRs are negligible, with only a few exceptions. Overall, both synthetic and simulated RIRs enhance system performance, with a slight advantage observed for synthetic RIRs. This suggests that while both approaches are beneficial, synthetic RIRs may offer a marginally better improvement.

We conclude that leveraging a large-scale synthetic or simulated RIR dataset for augmentation can significantly enhance the robustness and resilience of fake speech detection systems, mitigating the need for extensive real-world RIR collections. Notably,  \texttt{BR2}, trained with the large-scale synthetic RIRs, achieves outstanding performance, yielding the best results across all evaluation sets.

\section{Discussion}
\label{sec:discussion}

\begin{figure*}[t]
    \begin{subfigure}[b]{0.3\textwidth}
         \centering
         \includegraphics[width=\textwidth]{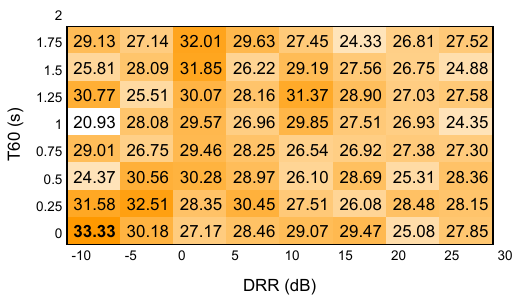} \\
         \includegraphics[width=\textwidth]{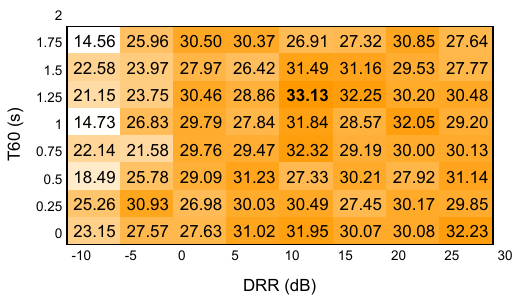} \\
         \includegraphics[width=\textwidth]{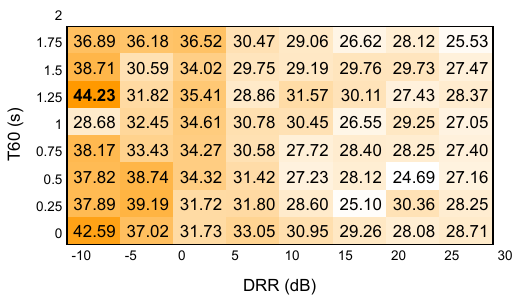}
         \caption{RawNet2}
     \end{subfigure}
     \hfill
     \begin{subfigure}[b]{0.3\textwidth}
         \centering
         \includegraphics[width=\textwidth]{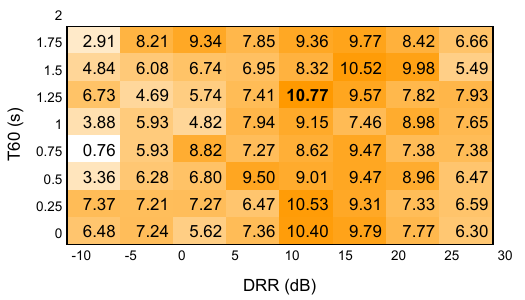} \\
         \includegraphics[width=\textwidth]{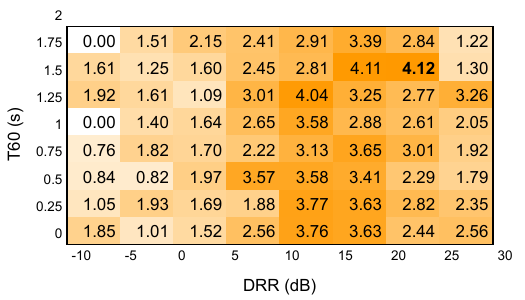}  \\
         \includegraphics[width=\textwidth]{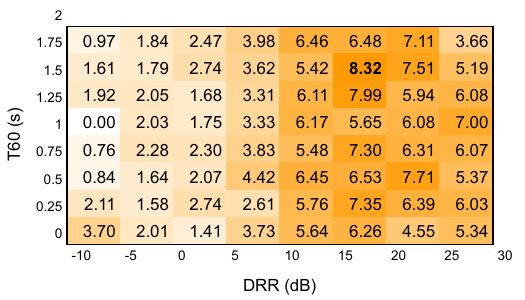}   \caption{Wav2Vec2 + AASIST}
     \end{subfigure}
     \hfill
     \begin{subfigure}[b]{0.3\textwidth}
         \centering
         \includegraphics[width=\textwidth]{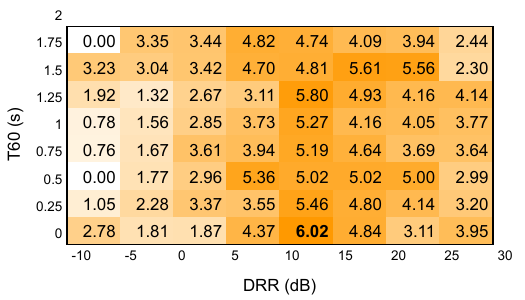} \\
         \includegraphics[width=\textwidth]{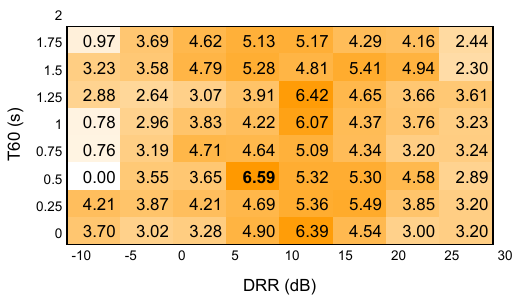}   \\
         \includegraphics[width=\textwidth]{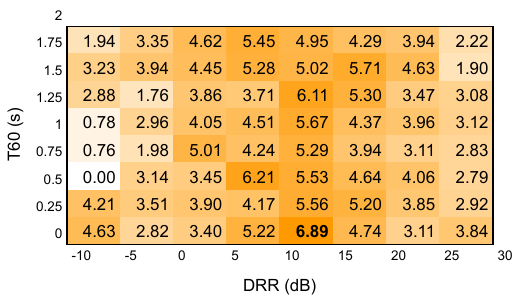} 
         \caption{Wav2Vec2 + Conformer}
     \end{subfigure}
     \vspace{-2mm}
        \caption{FAR (\%) of the systems trained with recorded RIRs (top row), synthetic RIRs (middle row), and simulated RIRs (bottom row) at the EER threshold test on the C1R2 set, results are split by T60 and DRR values of augmented RIRs.}
        \label{fig:far-propose}
  \centering
  \vspace{-2mm}
\end{figure*}

\subsection{Vulnerability to reverberant fake speech}
In addition to EER, we also evaluated the FAR of each system at the EER threshold using reverberant fake speech created with synthetic RIRs. Figure \ref{fig:far-baseline} shows the FAR of the three baseline systems tested on C1R2's subsets, categorized by the DRR and T60 values of the augmented RIRs.
Interestingly, \texttt{System A} exhibits a distinct pattern, where FAR increases significantly as DRR decreases. Conversely, \texttt{System B} lacks a clear pattern, yet reveals heightened sensitivity to specific combinations of DRR and T60 values.
\texttt{System C} has a more balanced result across both DRR and T60 axes, showing less variation in performance.
Notably, the absence of RIRs in the training process renders some detection systems more susceptible to specific types of reverberant fake speech, potentially exploitable by attackers to enhance their success rates.

\subsection{Overfitting to a small amount of RIRs}
Training with RIR augmentation generally improves performance, but may lead to overfitting if the amount of augmented RIRs is insufficient. Notably, \texttt{BR1} consistently underperforms compared to \texttt{B} across all evaluation sets.
To further evaluate the performance of \texttt{BR1}, we conducted additional tests using the second-best checkpoint from the 17th epoch, as the best model was previously determined to be the 47th epoch checkpoint.
The EERs for the 17th epoch checkpoint is 2.15\% (C1), 2.95\% (Pooled), 6.13\% (C1R1), and 5.10\% (C1R2), representing a slight improvement over the 47th epoch result (\texttt{BR1}, Table \ref{tab:eer}), yet still inferior to \texttt{System B}'s performance on reverberant fake speech.
We suspect that \texttt{BR1} overfitted to the limited set of 234 RIRs used for augmentation, resulting in degraded performance.

\subsection{Robust fake speech speech detection systems}

Figure \ref{fig:far-propose} presents a comparison of FARs for systems trained with different types of RIRs: recorded RIRs (top row), synthetic RIRs (middle row), and simulated RIRs (bottom row).
The overall FARs exhibit better balance compared to the baseline systems (Fig. \ref{fig:far-baseline}).
Interestingly, systems trained with simulated RIRs exhibit slightly increased vulnerability to specific RIR conditions, a trend most pronounced in the case of \texttt{AR3}.
This outcome is unsurprising, as the simulated RIRs were designed to mimic specific room sizes rather than uniformly capture the T60 and DRR characteristics.

Furthermore, systems trained with RIR augmentation demonstrate improved performance on the original DF evaluation set, which encompasses various coding and compression conditions, without the addition of augmentation methods targeting these conditions.
We argue that generalization is crucial for fake speech detection systems, as speech utterances can be manipulated in numerous unpredictable ways \cite{muller2024new}, many of which will be unforeseen at deployment time.
Boosting training data variety helps prevent the system from taking shortcuts \cite{geirhos2020shortcut,shim2023construct}, ensuring more robust solutions, and enables systems to detect novel manipulation methods and adapt to evolving threats.

\section{Conclusion}
\label{sec:conclusion}

In this paper, we explored the potential of using RIRs to evade automatic fake speech detection systems. Our findings revealed that RIR can significantly degrade the performance of SOTA models, doubling their EER.
Although large-scale RIR datasets can enhance model robustness, their scarcity makes synthetic RIRs an attractive alternative, providing a cost-effective and scalable means of obtaining high-quality training data.
Our experiments demonstrated that augmenting with synthetic RIRs significantly enhances model generalization. Notably, our top-performing system achieved substantial EER reductions: from 2.58\% to 2.13\% on the DF task and from 4.00\% to 2.53\% on the reverberant fake speech task.
This approach is model-agnostic, making it versatile and widely applicable.
Furthermore, our experiments suggest that while a large-scale simulated RIR dataset can be beneficial, a synthetic RIR dataset yields even better results.
For deep fake detection, robustness is more crucial than benchmark EER, as it must detect new synthesized speech types and evolving audio manipulations.

%\section{ACKNOWLEDGMENTS}
%\label{sec:ack}

\bibliographystyle{IEEEbib}
\bibliography{refs}

\begin{thebibliography}{10}

\bibitem{floridi2020gpt}
Luciano Floridi and Massimo Chiriatti,
\newblock ``Gpt-3: Its nature, scope, limits, and consequences,''
\newblock {\em Minds and Machines}, vol. 30, pp. 681--694, 2020.

\bibitem{rombach2022high}
Robin Rombach, Andreas Blattmann, Dominik Lorenz, Patrick Esser, and Bj{\"o}rn Ommer,
\newblock ``High-resolution image synthesis with latent diffusion models,''
\newblock in {\em Proceedings of the IEEE/CVF conference on computer vision and pattern recognition}, 2022, pp. 10684--10695.

\bibitem{thies2016face2face}
Justus Thies, Michael Zollhofer, Marc Stamminger, Christian Theobalt, and Matthias Nie{\ss}ner,
\newblock ``Face2face: Real-time face capture and reenactment of rgb videos,''
\newblock in {\em Proceedings of the IEEE conference on computer vision and pattern recognition}, 2016, pp. 2387--2395.

\bibitem{chan2019everybody}
Caroline Chan, Shiry Ginosar, Tinghui Zhou, and Alexei~A Efros,
\newblock ``Everybody dance now,''
\newblock in {\em Proceedings of the IEEE/CVF international conference on computer vision}, 2019, pp. 5933--5942.

\bibitem{kreuk2022audiogen}
Felix Kreuk, Gabriel Synnaeve, Adam Polyak, Uriel Singer, Alexandre D{\'e}fossez, Jade Copet, Devi Parikh, Yaniv Taigman, and Yossi Adi,
\newblock ``Audiogen: Textually guided audio generation,''
\newblock {\em arXiv preprint arXiv:2209.15352}, 2022.

\bibitem{arik2018neural}
Sercan Arik, Jitong Chen, Kainan Peng, Wei Ping, and Yanqi Zhou,
\newblock ``Neural voice cloning with a few samples,''
\newblock in {\em Proc. NeurIPS}, 2018, vol.~31.

\bibitem{masood2023deepfakes}
Momina Masood, Mariam Nawaz, Khalid~Mahmood Malik, Ali Javed, Aun Irtaza, and Hafiz Malik,
\newblock ``Deepfakes generation and detection: State-of-the-art, open challenges, countermeasures, and way forward,''
\newblock {\em Applied intelligence}, vol. 53, no. 4, pp. 3974--4026, 2023.

\bibitem{kinnunen2017asvspoof}
Tomi Kinnunen, Md~Sahidullah, H{\'e}ctor Delgado, Massimiliano Todisco, Nicholas Evans, Junichi Yamagishi, and Kong~Aik Lee,
\newblock ``The asvspoof 2017 challenge: Assessing the limits of replay spoofing attack detection,''
\newblock in {\em Proc. Interspeech}, 2017.

\bibitem{todisco2019asvspoof}
Massimiliano Todisco, Xin Wang, Ville Vestman, Md. Sahidullah, Héctor Delgado, Andreas Nautsch, Junichi Yamagishi, Nicholas Evans, Tomi~H. Kinnunen, and Kong~Aik Lee,
\newblock ``Asvspoof 2019: Future horizons in spoofed and fake audio detection,''
\newblock in {\em Proc. INTERSPEECH}, 2019, pp. 1008--1012.

\bibitem{yamagishi2021asvspoof}
Junichi Yamagishi, Xin Wang, Massimiliano Todisco, Md~Sahidullah, Jose Patino, Andreas Nautsch, Xuechen Liu, Kong~Aik Lee, Tomi Kinnunen, Nicholas Evans, and Héctor Delgado,
\newblock ``Asvspoof 2021: accelerating progress in spoofed and deepfake speech detection,''
\newblock in {\em Proc. ASVspoof Challenge}, 2021, pp. 47--54.

\bibitem{casanova2022yourtts}
Edresson Casanova, Julian Weber, Christopher~D Shulby, Arnaldo~Candido Junior, Eren G{\"o}lge, and Moacir~A Ponti,
\newblock ``Yourtts: Towards zero-shot multi-speaker tts and zero-shot voice conversion for everyone,''
\newblock in {\em International Conference on Machine Learning}. PMLR, 2022, pp. 2709--2720.

\bibitem{shen2023naturalspeech}
Kai Shen, Zeqian Ju, Xu~Tan, Yanqing Liu, Yichong Leng, Lei He, Tao Qin, Sheng Zhao, and Jiang Bian,
\newblock ``Naturalspeech 2: Latent diffusion models are natural and zero-shot speech and singing synthesizers,''
\newblock {\em arXiv preprint arXiv:2304.09116}, 2023.

\bibitem{vyas2023audiobox}
Apoorv Vyas, Bowen Shi, Matthew Le, Andros Tjandra, Yi-Chiao Wu, Baishan Guo, Jiemin Zhang, Xinyue Zhang, Robert Adkins, William Ngan, et~al.,
\newblock ``Audiobox: Unified audio generation with natural language prompts,''
\newblock {\em arXiv preprint arXiv:2312.15821}, 2023.

\bibitem{muller2024new}
Nicolas~M M{\"u}ller, Piotr Kawa, Shen Hu, Matthias Neu, Jennifer Williams, Philip Sperl, and Konstantin B{\"o}ttinger,
\newblock ``A new approach to voice authenticity,''
\newblock {\em arXiv preprint arXiv:2402.06304}, 2024.

\bibitem{steinmetz2021filtered}
Christian~J Steinmetz, Vamsi~Krishna Ithapu, and Paul Calamia,
\newblock ``Filtered noise shaping for time domain room impulse response estimation from reverberant speech,''
\newblock in {\em Proc. WASPAA}. IEEE, 2021, pp. 221--225.

\bibitem{singh2021image2reverb}
Nikhil Singh, Jeff Mentch, Jerry Ng, Matthew Beveridge, and Iddo Drori,
\newblock ``Image2reverb: Cross-modal reverb impulse response synthesis,''
\newblock in {\em Proceedings of the IEEE/CVF International Conference on Computer Vision}, 2021, pp. 286--295.

\bibitem{ratnarajah2023av}
Anton Ratnarajah, Sreyan Ghosh, Sonal Kumar, Purva Chiniya, and Dinesh Manocha,
\newblock ``Av-rir: Audio-visual room impulse response estimation,''
\newblock {\em arXiv preprint arXiv:2312.00834}, 2023.

\bibitem{ko2017study}
Tom Ko, Vijayaditya Peddinti, Daniel Povey, Michael~L Seltzer, and Sanjeev Khudanpur,
\newblock ``A study on data augmentation of reverberant speech for robust speech recognition,''
\newblock in {\em Proc. ICASSP)}. IEEE, 2017, pp. 5220--5224.

\bibitem{salamon2017deep}
Justin Salamon and Juan~Pablo Bello,
\newblock ``Deep convolutional neural networks and data augmentation for environmental sound classification,''
\newblock {\em IEEE Signal processing letters}, vol. 24, no. 3, pp. 279--283, 2017.

\bibitem{bryan2020impulse}
Nicholas~J Bryan,
\newblock ``Impulse response data augmentation and deep neural networks for blind room acoustic parameter estimation,''
\newblock in {\em Proc. ICASSP}, 2020, pp. 396--400.

\bibitem{traer2016statistics}
James Traer and Josh~H McDermott,
\newblock ``Statistics of natural reverberation enable perceptual separation of sound and space,''
\newblock {\em PNAS}, vol. 113, no. 48, pp. E7856--E7865, 2016.

\bibitem{allen1979image}
Jont~B Allen and David~A Berkley,
\newblock ``Image method for efficiently simulating small-room acoustics,''
\newblock {\em The Journal of the Acoustical Society of America}, vol. 65, no. 4, pp. 943--950, 1979.

\bibitem{carlini2018audio}
Nicholas Carlini and David Wagner,
\newblock ``Audio adversarial examples: Targeted attacks on speech-to-text,''
\newblock in {\em 2018 IEEE security and privacy workshops (SPW)}. IEEE, 2018, pp. 1--7.

\bibitem{xie2021enabling}
Yi~Xie, Zhuohang Li, Cong Shi, Jian Liu, Yingying Chen, and Bo~Yuan,
\newblock ``Enabling fast and universal audio adversarial attack using generative model,''
\newblock in {\em Proceedings of the AAAI conference on artificial intelligence}, 2021, vol.~35, pp. 14129--14137.

\bibitem{chen2021pindrop}
Tianxiang Chen, Elie Khoury, Kedar Phatak, and Ganesh Sivaraman,
\newblock ``Pindrop labs’ submission to the asvspoof 2021 challenge,''
\newblock in {\em Proc. ASVspoof 2021 Challenge}, 2021, pp. 89--93.

\bibitem{tomilov21asvspoof}
Anton Tomilov, Aleksei Svishchev, Marina Volkova, Artem Chirkovskiy, Alexander Kondratev, and Galina Lavrentyeva,
\newblock ``Stc antispoofing systems for the asvspoof2021 challenge,''
\newblock in {\em Proc. ASVspoof 2021 Challenge}, 2021, pp. 61--67.

\bibitem{tak2021endtoend}
Hemlata Tak, Jose Patino, Massimiliano Todisco, Andreas Nautsch, Nicholas Evans, and Anthony Larcher,
\newblock ``End-to-end anti-spoofing with rawnet2,''
\newblock in {\em Proc. ICASSP}, 2021, pp. 6369--6373.

\bibitem{tak2022rawboost}
Hemlata Tak, Madhu Kamble, Jose Patino, Massimiliano Todisco, and Nicholas Evans,
\newblock ``Rawboost: A raw data boosting and augmentation method applied to automatic speaker verification anti-spoofing,''
\newblock in {\em Proc. ICASSP}, 2022, pp. 6382--6386.

\bibitem{tak2022automatic}
Hemlata Tak, Massimiliano Todisco, Xin Wang, Jee weon Jung, Junichi Yamagishi, and Nicholas Evans,
\newblock ``Automatic speaker verification spoofing and deepfake detection using wav2vec 2.0 and data augmentation,''
\newblock in {\em Proc. Odyssey}, 2022, pp. 112--119.

\bibitem{baevski2020wav2vec}
Alexei Baevski, Yuhao Zhou, Abdelrahman Mohamed, and Michael Auli,
\newblock ``wav2vec 2.0: A framework for self-supervised learning of speech representations,''
\newblock {\em NeurIPS}, vol. 33, pp. 12449--12460, 2020.

\bibitem{jung2022aasist}
Jee-weon Jung, Hee-Soo Heo, Hemlata Tak, Hye-jin Shim, Joon~Son Chung, Bong-Jin Lee, Ha-Jin Yu, and Nicholas Evans,
\newblock ``Aasist: Audio anti-spoofing using integrated spectro-temporal graph attention networks,''
\newblock in {\em ICASSP}, 2022, pp. 6367--6371.

\bibitem{rosello2023conformer}
Eros Rosello, Alejandro Gomez-Alanis, Angel~M. Gomez, and Antonio Peinado,
\newblock ``A conformer-based classifier for variable-length utterance processing in anti-spoofing,''
\newblock in {\em Proc. INTERSPEECH}, 2023, pp. 5281--5285.

\bibitem{gulati2020conformer}
Anmol Gulati, James Qin, Chung-Cheng Chiu, Niki Parmar, Yu~Zhang, Jiahui Yu, Wei Han, Shibo Wang, Zhengdong Zhang, Yonghui Wu, and Ruoming Pang,
\newblock ``Conformer: Convolution-augmented transformer for speech recognition,''
\newblock in {\em INTERSPEECH}, 2020, pp. 5036--5040.

\bibitem{geirhos2020shortcut}
Robert Geirhos, J{\"o}rn-Henrik Jacobsen, Claudio Michaelis, Richard Zemel, Wieland Brendel, Matthias Bethge, and Felix~A Wichmann,
\newblock ``Shortcut learning in deep neural networks,''
\newblock {\em Nature Machine Intelligence}, vol. 2, no. 11, pp. 665--673, 2020.

\bibitem{shim2023construct}
Hye-jin Shim, Rosa~Gonz{\'a}lez Hautam{\"a}ki, Md~Sahidullah, and Tomi Kinnunen,
\newblock ``How to construct perfect and worse-than-coin-flip spoofing countermeasures: A word of warning on shortcut learning,''
\newblock {\em arXiv preprint arXiv:2306.00044}, 2023.

\end{thebibliography}

\end{document}